# 2D MXenes: Visible Black but Infrared White Materials

***Yang Li*[1,3]*, Cheng Xiong*[1,3]*, He Huang*[1,3]*, Xudong Peng*[1]*, Deqing Mei*[2]*, Meng Li*[1]*, Gongze Liu*[1]*, Maochun Wu*[1]*, Tianshou Zhao*[1]** *& Baoling Huang*[1]***

## Abstract

Black materials with low infrared absorption/emission (or IR white) are rare in nature but highly desired in numerous areas, such as solar-thermal energy harvesting, multispectral camouflage, thermal insulation, and anti-counterfeiting. Due to the lack of spectral selectivity in intrinsic materials, such counter-intuitive properties are generally realized by constructing complicated subwavelength metamaterials with costly nanofabrication techniques. Here we report the low mid-IR emissivity (down to 10%) of 2D $Ti_3C_2T_x$ MXenes. Associated with a high solar absorptance (up to 90%), they embrace the best spectral selectivity among the reported intrinsic black solar absorbing materials. Their appealing potentials in several aforementioned areas are experimentally demonstrated. First-principles calculations reveal that the IR emissivity of MXenes relies on both the nanoflake orientations and terminal groups, indicating great tunability. The calculations also suggest that more MXenes including $Ti_2CT_x$, $Nb_2CT_x$, and $V_2CT_x$ are also potential low-emissivity materials. This work opens the avenue to further exploration of a family of intrinsically low-emissivity materials with over 70 members.

Keywords: MXenes, $Ti_3C_2T_x$, infrared emissivity, solar-thermal, camouflage, anti-counterfeiting

## 1. Introduction

Visible black materials with intrinsically low mid-IR emissivity (IR white) are scarce but of great significance for many areas, such as solar energy harvesting,[1,2] building thermal insulation,[3] multispectral camouflage,[4,5] and container package[6]. For instance, in solar-thermal energy conversion systems, to maximize the conversion efficiency, ideal solar absorbing materials should show perfect absorption over the broad solar spectrum (0.3-2.5 μm), but simultaneously ultralow mid-IR absorption/emission (2.5-20 μm) to suppress enormous heat dissipation via IR thermal re-radiation (e.g. 0.68 kW $m^{-2}$ @ 100 °C, 0.68 sun), known as selective solar absorption.[1,7] Intuitively, under the same solar irradiance, the low-emissivity absorbing materials reach much higher temperatures than those blackbody materials, such as carbon-based materials,[8,9] black polymers,[10] and black oxides.[11,12] The more generated thermal energy and larger temperature rises are highly beneficial for a lot of heat-driven applications including passive heating,[13] anti-icing/deicing,[14] steam generation,[1] catalysis,[15] actuators,[16] and electricity generation.[17] Moreover, even with higher surface temperatures, their low emissivity endows black materials with the counter-intuitive ability to conceal themselves from IR detection due to their limited thermal radiation, showing great potential in multispectral (IR and visible) camouflage assisted by their dark colors.[4,18]

[1] Department of Mechanical and Aerospace Engineering, The Hong Kong University of Science and Technology, Clear Water Bay, Kowloon, Hong Kong 999077, China
[2] School of Mechanical Engineering, Zhejiang University, Hangzhou 310027, China
[3] These authors contributed equally.
*Corresponding author. Email address: mebhuang@ust.hk; metzhao@ust.hk

However, the lack of ideal visible black but IR white intrinsic materials has been a long-standing challenge for decades.[17] For this reason, great efforts have been made in designing subwavelength metamaterials or metasurfaces through various light manipulation strategies, including photonic crystals,[19,20] plasmon resonances,[2,7,21] and interference effects.[22-24] However, there are at least two limitations in these artificial low-emissivity metamaterials. Most of these selective metamaterials are pore-free dense structures with rigid substrates, which significantly hinder the diversity of their application scenarios. In addition, the properties of these subwavelength metamaterials are extremely sensitive to the feature size of nanostructures, requiring precise lithography or high-vacuum deposition techniques, which inherently restricts their cost reduction. It is preferable if there was a flexible intrinsic material with strong, broadband solar absorption but low IR emissivity. In this context, we resorted to a new family of two-dimensional (2D) materials, MXenes, composed of transitional metal carbides and nitrides. Since they were first reported in 2011,[25] MXenes have brought substantial benefits and new opportunities to a variety of areas ranging from energy storage,[26,27] and catalysis[28] to electromagnetic wave management.[29-33] Specifically, the outstanding ability of $Ti_3C_2T_x$ (T denotes the -O-, -OH, -F and other terminal groups) MXenes in electromagnetic wave absorption has enabled both the electromagnetic interference shield effects (in the microwave region)[29,30] and photothermal effects (in the visible and near-IR regions).[31-33] However, the interaction between the MXenes and electromagnetic waves in the mid-IR region (>2.5 μm), which dominates the radiative heat transfer near room temperature, is rarely explored.

In this study, for the first time, we discovered that unlike the strong absorption in visible, near-IR, and microwave regions, the black $Ti_3C_2T_x$ MXene films show strong reflection up to 90% for the mid-IR wavelengths, resulting in a quite low emissivity. Associated with its high absorption (90%) across the solar spectrum, it achieves a record-high spectral selectivity (8.2) among reported intrinsic solar absorbing materials, and thus the highest solar-thermal conversion efficiency (78% under 1 sun, at 100 °C), to the best of our knowledge. Excellent intrinsic spectral selectivity is offered by $Ti_3C_2T_x$ MXene in a free-standing form as well as coatings on versatile substrates, even on porous rough surfaces, showing much wider applications than metamaterial selective absorbers. Further, first-principles calculations reveal that the optical properties of the $Ti_3C_2T_x$ MXenes are potentially tunable due to the dependence on the terminal groups and the orientation of nanoflakes. Other commercially available MXenes such as $Ti_2CT_x$, $Nb_2CT_x$, and $V_2CT_x$ are also potential infrared white materials.

## 2. Results and discussion

### 2.1. Low Mid-IR Emissivity of $Ti_3C_2T_x$ MXenes

As widely acknowledged, polished metals generally possess high light reflection over an ultra-broad band due to their high damping constant and low refractive index (**Figure 1**a). For instance, as shown in Figure 1b, the stainless steel (SLS) has both a low solar absorptance $\bar{\alpha}$ of 38% and a low IR emissivity $\bar{\varepsilon}$ of 9% at 100 °C. In contrast, most of the black materials such as carbon-based absorbers have strong UV-visible-IR absorption (Figure 1c). A black absorber made of carbon nanotubes (CNT) shows not only a $\bar{\alpha}$ as high as 95%, but also a near-unity $\bar{\varepsilon}$ of 93% (Figure 1d). The blackbody-like absorber is not an ideal solar absorber, since most of the harvested solar energy will be dissipated via the massive thermal re-radiation (at 100 °C), leading to a reduced solar-thermal efficiency $\eta_{solar\text{-}th}$ of ~32% under the illumination of 1 sun (1 kW $m^{-2}$) near room temperature.

Unlike metals and blackbodies, a free-standing 15-μm-thick $Ti_3C_2T_x$ MXene film exhibits a high $\bar{\alpha}$ of 90% comparable to that of the CNT absorber, while beyond the solar spectrum, its absorption rapidly declines to a rather low level due to the increasing mid-IR reflection (Figure 1e,f). Its $\bar{\varepsilon}$ is only around 17% at 100 °C, suggesting excellent spectral selectivity ($\bar{\alpha}/\bar{\varepsilon}$). Moreover, the $\bar{\varepsilon}$ of the $Ti_3C_2T_x$ film can be further reduced to as low as 10% by controlling the surface morphology as discussed later, which is even close to that of some polished metals (9% for W and SLS) (Figure 1h). To visually demonstrate the low $\bar{\varepsilon}$ of the $Ti_3C_2T_x$ film, it was placed on a hot plate with a constant temperature of 100 °C, and characterized by an IR imager (the default emissivity was set as 95% in this work). Despite the identical real temperature of 100 °C, the $Ti_3C_2T_x$ film appears much colder than the CNT absorber in the IR image, and as cold as the polished SLS with a comparably low $\bar{\varepsilon}$ (Figure 1i). The $\eta_{solar\text{-}th}$ (1 sun, 100 °C) of the low-emissivity $Ti_3C_2T_x$ film can reach a record-high value of 78% for intrinsic black materials, far higher than that of the CNT absorber (32%). To the best of our knowledge, both the $\bar{\alpha}/\bar{\varepsilon}$ and $\eta_{solar\text{-}th}$ of the $Ti_3C_2T_x$ MXene are the highest among all the intrinsic solar absorbing materials ($\bar{\alpha}$ > 50%) reported so far (Table S1). By virtue of the higher $\eta_{solar\text{-}th}$, its measured surface temperature rise (62 °C) is higher than that of the CNT absorber (50 °C) under 1-sun illumination in the open air (Figure 1g).

### 2.2. Fabrication and Characterizations of $Ti_3C_2T_x$ MXene Films

The free-standing black $Ti_3C_2T_x$ MXene film with great flexibility in this work was prepared by a standard process reported previously,[27,29,34] which includes chemically etching of the Al atoms from $Ti_3AlC_2$ phases, nanoflakes delamination by centrifugation, and vacuum-assisted filtration. The TEM image of the prepared $Ti_3C_2T_x$ powder after delamination verifies that 2D nanoflakes (>1 μm in lateral size) with few layers were attained and the thickness of each layer was around 1 nm, agreeing with the results in the literature (Figure S1).[34] The cross-sectional SEM image of the filtrated film shows that it has a 15 μm thickness and is stacked by well-aligned nanoflakes (**Figure 2**a). Both the X-ray diffraction (XRD) and the X-ray photoelectron spectroscopy (XPS) were performed to characterize the vacuum-filtrated $Ti_3C_2T_x$ film (the top side). The XRD pattern with a pronounced peak at $2\theta = 6.5°$ is assigned to the (002) plane of the $Ti_3C_2T_x$ MXene (Figure 2b), which is well consistent with the results in the previous works.[29,35] Moreover, the XPS results verify the presence of Ti, C, O, and F in the film, with a percentage atomic composition of 22.8%, 42.4%, 26.3%, and 8.5%, respectively (Figure S2). From the high-resolution XPS spectra of O, F, Ti, and C, three typical types of terminal groups are identified in the $Ti_3C_2T_x$ film including oxide (-O-), hydroxyl (-OH), and fluoride (-F).

The absorption/emission properties of the prepared $Ti_3C_2T_x$ film over a wide wavelength range (0.3-16.7 μm) were evaluated using UV/visible/near-IR and Fourier transform infrared (FTIR) spectrometers equipped with integrating spheres. As shown in Figure 2c, both the top and bottom surfaces of the $Ti_3C_2T_x$ film show strong light absorption for the wavelengths from 0.3 to 1.2 μm and dramatically reduced absorption/emission for longer wavelengths. The absorption/emission features of the film can be maintained without notable decay after long-term (120 hours) thermal annealing in both the vacuum (< $7\times10^{-2}$ Torr) at 400 °C and the ambient air at 200 °C (Figure S3). To figure out this transition in the absorption spectra, the optical properties (i.e., permittivity $\varepsilon$, and refractive index $n$) of the film were measured using spectroscopic ellipsometry. Over the visible and near-IR range, the real part of the permittivity ($\varepsilon_1$) exhibits an optical dispersion behavior with a transition from positive to negative as the wavelength increases, indicating a change from dielectric to metallic response at longer wavelengths (Figure S4). This transition to metallic behavior at 1.3 μm can well explain the increasing reflection beyond the transition wavelength, which is highly desirable to achieve low mid-IR emissivity.

Interestingly, the emissivity of the top surface is slightly higher than that of the bottom over the whole spectral range. The $\bar{\alpha}$ and $\bar{\varepsilon}$ @100 °C of the top are 90 ± 2% and 17 ± 2%, while those of the bottom are 82 ± 1% and 10 ± 1%, respectively. As a result, although with the same surface temperature (~100 °C), the bottom (39 °C) appears a bit colder than the top (45 °C) in the IR image (Figure 2d). The surface SEM images reveal that the bottom surface that adhered to the filter membrane during filtration is flatter than the top (Figure 2e). Their 3D morphologies given by a surface profiler also quantitatively verify the roughness of the bottom surface ($R_a$ = 215 nm) is around half of that for the top surface (410 nm) (Figure S5 and S6). Although the larger roughness of the top surface may contribute to the increase in solar absorption, it is is still too small in scale to have strong interaction with long-wavelength mid-IR light (~10 μm),[36] implying that the discrepancy in IR emissivity of the two surfaces is likely caused by factors lying behind the surface roughness.

During vacuum-assisted filtration, the boundary conditions of 2D $Ti_3C_2T_x$ nanoflakes near the two surfaces were quite different. The stacking and alignment of nanoflakes near the bottom was assisted by the flat filter membrane and strong vacuum pressure, while the top surface was formed under weaker constraints. In other words, a portion of the nanoflakes away from the filter membrane probably cannot be aligned parallel to the membrane during filtration, accumulatively forming a corrugated surface morphology on top, as evidenced in Figure 2e. Due to the structural anisotropy,[37] the in-plane and cross-plane optical properties (i.e., refractive index and permittivity) of 2D $Ti_3C_2T_x$ nanoflakes differ from each other. Therefore, the distinctions in the nanoflakes orientation near the two surfaces would result in emissivity variations. This hypothesis is proved by the first-principles calculations, as discussed later. In addition, Figure 2f shows the absorption/emission spectra of the $Ti_3C_2T_x$ films (bottom side) with varying thicknesses including 2, 5, 10, and 15 μm. Clearly, the emissivity of the optically thick films is independent on the film thickness in the thickness range considered.

### 2.3. Potential Applications of Low-Emissivity MXenes

Porous structures are indispensable components to a lot of solar-thermal applications, such as steam generation,[1] seawater desalination,[31] and smart textiles.[38] However, current high-performance selective solar absorbers are usually realized by constructing artificial metamaterials/metasurfaces in nanoscale (multilayer nanofilms or nanophotonic structures) on dense flat substrates,[7,20,22,39] which will lose their effectiveness when coated on porous substrates owing to the break of resonant conditions. For instance, we coated a previously reported high-performance metamaterial absorber (Ag film/TiN nanoparticles/$SiO_2$) on a highly porous Nylon 66 membrane (5 cm in diameter). The highly porous structure of the pristine Nylon 66 membrane is shown in the inset of **Figure 3**a. Despite the optimized film thicknesses, the metamaterial absorber only offers a low $\bar{\alpha}$ of 45%, much lower than the values (~95%) on dense substrates.[7] In contrast, the $Ti_3C_2T_x$ film (2 mg in weight) sustained its intrinsically high $\bar{\alpha}$ of 85% and low $\bar{\varepsilon}$ of 25% even coated on highly porous membranes (Figure 3a), because its optical properties are dominated by the material itself instead of the subwavelength microstructures. As a result, the $\eta_{solar\text{-}th}$ of the $Ti_3C_2T_x$-Nylon 66 absorber (68% under 1 sun, at 100 °C) is superior to that of its metamaterial-based counterpart (42%). The bending test shows that the $Ti_3C_2T_x$-Nylon 66 absorber can maintain its optical performance without notable decay after bending 150° for 10,000 times (Figure S7).

As shown in Figure 3b, under the exposure of 1 sun, the $Ti_3C_2T_x$-Nylon 66 was rapidly heated to 70 °C from 22 °C within 1 min. After 10 min exposure, the surface temperature reached saturation of around 89 °C, while those of the metamaterial-based absorber and the black paint were only 72 and 80 °C due to the lower $\eta_{solar\text{-}th}$. In short, intrinsic MXene absorbers are compatible with both dense and porous substrates, and can also work in the form of a free-standing film, which significantly expands application scenarios of selective solar absorbers. Moreover, unlike metamaterials, there is no need to employ time-consuming, expensive nanofabrication techniques for the preparation of such intrinsic absorbers, which make them more attractive in large-scale applications.

Besides the solar-thermal conversion, the low-emissivity black MXenes can benefit many other areas such as multispectral camouflage and anti-counterfeiting. In conventional IR camouflage coatings that are mostly made of metallic micro powders such as aluminum, both the high glossiness and high brightness limit their compatibility with visible and even near-IR light.[18,40] Figure 3c,d show the optical and IR images of a person who wore a white T-shirt with black $Ti_3C_2T_x$ coating ($0.16\times0.128$ m$^2$) at night. The temperature reading of the area with $Ti_3C_2T_x$ coating in the IR image was only 25 °C and much lower than the body temperature (~36 °C), appearing nearly as cold as the environment (23 °C). This result demonstrates that low-emissivity $Ti_3C_2T_x$ could be implemented as IR camouflage coatings to conceal human bodies from IR detection. Meanwhile, as shown in Figure 3c, the black $Ti_3C_2T_x$ coating on the white T-shirt was invisible at night, allowing the covered objects to blend in with the dark environment. In other words, the low-emissivity black $Ti_3C_2T_x$ can overcome the issues of conventional IR camouflage coatings and be an appealing alternative in both IR and visible camouflage.

Counterfeiting is causing tremendous losses in both the security and property of customers, companies, and governments. Anti-counterfeiting features are increasingly demanded to prevent valuable items such as brands, tickets, quick response (QR) codes, banknotes, and confidential documents from being replicated. Currently, most anti-counterfeiting technologies are enabled by photoluminescence[41] and magnetic response.[42] Here we demonstrated an alternative technology against counterfeiters by using IR-metameric security inks composed of low-emissivity MXenes. In Figure 3e, the black anti-counterfeiting word "HKUST" was written on a white (polyvinylidene fluoride) PVDF membrane using both the commercial black paint (for "HK") and the $Ti_3C_2T_x$ MXene solution (for "UST"), which cannot be told apart by naked eyes because of the similar spectral features in visible light. However, the "HK" almost disappeared under the IR thermal imager due to the small contrast in $\bar{\varepsilon}$ of the PVDF (~95%) and the black paint (~92%), while the "UST" can still be clearly observed at different set temperatures because of the much lower $\bar{\varepsilon}$ of the $Ti_3C_2T_x$ (Figure 3f,g).

### 2.4. First-Principles Calculations

To explore the underlying mechanisms of the spectral selectivity in the $Ti_3C_2T_x$, and search for more potential low-emissivity MXenes, we investigated their optical properties by the density functional theory (DFT) calculations. As shown in **Figure 4**a, bulk $Ti_3C_2T_x$ films made of layered flakes with and without different terminal groups (-OH, -O, and -F) under the normal incidence of light were calculated. Referring to the Fresnel equation, the surface light reflection (*R*) from materials in the air at normal incidence can be obtained by

$$r(\omega)=\left|\frac{\sqrt{\varepsilon_1(\omega)+i\varepsilon_2(\omega)}-1}{\sqrt{\varepsilon_1(\omega)+i\varepsilon_2(\omega)}+1}\right|^2 \tag{1}$$

where $\omega$ is the angular frequency and $\varepsilon(\omega)=\varepsilon_1(\omega)+i\varepsilon_2(\omega)$ is the permittivity. Then, for optically thick materials with no transmission, the light absorption can be expressed as

$$\alpha(\omega)=1-r(\omega) \tag{2}$$

To achieve low reflection and therefore high absorption, $\varepsilon_1$ and $\varepsilon_2$ should approach 1 and 0 ($\varepsilon_2 > 0$). On the contrary, large $\varepsilon_1$ or $\varepsilon_2$ values lead to high surface reflection, and therefore low absorption/emission.

It should be noted that at normal incidence (wave vector $\vec{k}$ along the z-direction), the electric component of light orients in the x-y plane (Figure 4a). Hence, it is the in-plane permittivity that dominates the reflection. Interestingly, for the four kinds of $Ti_3C_2T_x$, both the in-plane $\varepsilon_1$ and $\varepsilon_2$ are relatively smaller in the UV-visible-NIR range but increase to quite large levels in the mid-IR region (Figure 4b,c). As a result, they offer wavelength selectivity in light absorption as expected (Figure 4d). In particular, the in-plane $\varepsilon_1$ of both the $Ti_3C_2(OH)_2$ and the $Ti_3C_2F_2$ undergo a transition to negative values in the near-IR region (~0.9 µm for $Ti_3C_2(OH)_2$ and ~1.3 µm for $Ti_3C_2F_2$), suggesting a strong metallic response to IR light. The transition to negative permittivity is consistent with the measured results. As a result, both the $Ti_3C_2(OH)_2$ and the $Ti_3C_2F_2$ exhibit high solar absorption and low IR emission.

Moreover, the DFT results verify that the in-plane permittivity of the $Ti_3C_2T_x$ MXenes is different from that along the z-direction (Figure 4e,f). This well explains why the absorption properties of the films strongly depend on the orientations of nanoflakes. As shown in Figure 4g and S8, for $Ti_3C_2T_x$ with various terminal groups, their absorptance calculated from the in-plane permittivity shows much better spectral selectivity than that from the z-direction, indicating that well-aligned nanoflakes parallel to the substrate are preferable, which is consistent with the experiments. Due to the differences in composition, the absorption spectra of the $Ti_3C_2T_x$ given by the DFT calculations do not exactly agree with the experimental data. However, they validate the great spectral selectivity, and demonstrate the dependences on terminal groups and nanoflake orientations. Apart from the $Ti_3C_2T_x$, the DFT calculations suggest that other MXenes such as $Ti_2CT_x$, $Nb_2CT_x$, and $V_2CT_x$ are also promising low-emissivity black materials (Figure 4h and S9).

## 3. Conclusions

In summary, we for the first time report on a group of 2D MXenes as black materials with a low emissivity. It is demonstrated that the free-standing $Ti_3C_2T_x$ film prepared by vacuum-assisted filtration features both a high solar absorptance up to 90% and a low IR emissivity down to 10%, yielding the highest spectral selectivity for intrinsic solar absorbing materials reported so far. As a result, under 1 sun illumination in the open air, the MXene film achieves a high temperature rise around 62 °C with respect to the ambient air. By comparison, the temperature of the CNT absorber with a higher emissivity only increases by 50 °C, due to the higher photo-thermal conversion efficiency provided by the lower emissivity. First-principles calculations reveal that the absorption/emission properties of the MXenes strongly rely on the orientation of the nanoflakes, and the terminal groups. Specifically, lower emissivity is obtained in those films with nanoflakes well-aligned parallel to the substrates, and -OH and/or -F terminal groups. The highly selective, flexible, black MXenes show great potential in solar-thermal energy conversion, adaptive IR camouflage, thermal insulation, and anti-counterfeiting.

## 4. Experimental Section

**Fabrication of the MXene Films:** The $Ti_3C_2T_x$ colloidal solution was prepared by the liquid-phase delamination of $Ti_3AlC_2$ powder. In detail, 1.98 g of lithium fluoride (LiF) (Alfa Aesar, 98.5%) was added to 35 mL of 9 M HCl aqueous solution. Then, 2 g of $Ti_3AlC_2$ powder was added to the mixture. After etching for 24 h at 35 °C, the solution was washed and centrifuged with deionized water until the supernatant reached a pH value of 6. Next, to delaminate the MXene, 1 g etched MXene was dispersed into 0.5 L DI water, and deaerated with Ar, followed by sonication for 1 h. The mixture was then centrifuged for 1 h at 3500 rmp, and the supernatant with dark green color was collected. After that, a certain amount of MXene dispersion solution was vacuum filtrated through a hydrophilic Celgard 3501 membrane, and the obtained MXene coated membrane was further dried under 60 °C in the vacuum oven. After peeling off from the filter membrane, a free-standing $Ti_3C_2T_x$ film was obtained. The concentration of the solution was determined by testing the weight of the dried MXene film. Free-standing $Ti_3C_2T_x$ films of different thicknesses were prepared using the same process. In addition, thin $Ti_3C_2T_x$ films (~2 mg) were also coated on Nylon 66 membranes for demonstration using vacuum-assisted filtration.

**Material Characterizations:** The morphology of the $Ti_3C_2T_x$ nanoflakes was observed by transmission electron microscopy (TEM, JEM-2010F, Jeol). The phase identification of the MXene film was conducted by using an X-ray diffractometer (XRD, PANalytical) with Cu Kα radiation at 45 kV and 40 mA. The surface and cross-section morphologies were characterized by scanning electron microscopy (SEM, JSM-7100F, Jeol). The composition of the film was characterized by X-ray photoelectron spectroscopy (XPS, PHI 5600 multi-technique system, Physical Electronics). A 3D optical profiler (NPFLEX, Bruker) was used to characterize the morphology and roughness of the film. The surface roughness was obtained by analyzing 307,200 data points with a vertical resolution of 0.15 nm.

**Optical Measurements:** The UV-visible-NIR (0.3-2.5 μm) reflectance (*R*) and transmittance (*T*) spectra of the samples were measured using a spectrometer (Lambda 950, Perkin Elmer) equipped with a 150 mm integrating sphere. The mid-IR (MIR, > 2.5 μm) *R* and *T* spectra were measured using a Fourier transform infrared spectrometer (FTIR, Vertex 70, Bruker) with a gold integrating sphere. The absorptance (*A*) spectra were directly derived from 1-*R*-*T*.

**Thermal Measurements:** To measure the temperature rise under solar illumination, the MXene film and reference samples were attached to the surface of polystyrene foam. A solar simulator (Oriel Sol2A, Newport) using a xenon lamp was used to provide standard and stable 1-sun power (1 kW $m^{-2}$). T-type thermal couples, which were connected to a data acquisition device (NI9213, National Instrument), were attached on the backside of samples to measure the steady-state temperatures. The temperature data were recorded every two seconds. A thermal imager (Ti25, Fluke) was used to taken IR photos of the samples.

**Solar Absorptance and Thermal Emissivity Calculation:** The spectrally averaged solar absorptance $\bar{\alpha}$ is defined as[39]

$$\bar{\alpha} = \frac{\int_{0.3\mu m}^{4\mu m} \alpha(\lambda) E_{solar}(\lambda) d\lambda}{I_{solar}} \tag{3}$$

The spectrally averaged thermal emissivity $\bar{\varepsilon}$ at 100 °C is calculated by[39]

$$\bar{\varepsilon}(T) = \frac{\int_{0.3\mu m}^{20\mu m} \varepsilon(\lambda) E_b(\lambda, T) d\lambda}{\sigma T^4} \tag{4}$$

Here $E_{solar}(\lambda)$, $E_b(\lambda,T)$, $\alpha(\lambda)$, and $\varepsilon(\lambda)$ represent the spectral solar power (AM 1.5G), the blackbody emission at $T$ = 100 °C, the absorptance, and the emissivity at the wavelength $\lambda$, respectively. $I_{solar}$ is the total solar irradiance (AM 1.5G, 1 sun or 1 kW $m^{-2}$), and $\sigma$ is the Stefan-Boltzmann constant.

**Solar-Thermal Conversion Efficiency Calculation:** The solar-thermal energy conversion efficiency under 1 sun can be obtained by[39]

$$\eta_{solar-th}(T) = \bar{\alpha} - \bar{\varepsilon}\frac{\sigma(T^4 - T_0^{\,4})}{I_{solar}} \tag{5}$$

where $T$ is the operating temperature, and $T_0$ is the ambient temperature.

**Thermal stability tests:** Thermal annealing cycles (24 hours × 5 cycles) in ambient air at both 100 and 200 °C were conducted by placing the samples on a hot plate (KW-4AH, Chemat Technology Inc.). Thermal annealing cycles (24 hours × 5 cycles) in vacuum at both 300 and 400 °C were conducted in a quartz tube furnace (OTF-1200X, MTI Corporation). The samples placed in a ceramic crucible were put in the quartz tube and sealed. The sealed quartz tube was evacuated to reach a vacuum atmosphere (<$7\times10^{-2}$ Torr). Afterward, the samples were heated up to 300 °C (or 400 °C) at a rate of 10 °C/min and annealed for 24 h $cycle^{-1}$.

**Supporting Information**

Supporting Information is available from the author.

## Acknowledgments

The authors would like to acknowledge for the financial support from the Hong Kong General Research Fund (Grant No. 16214217). D.M. and Y.L. also acknowledge the financial support from the National Natural Science Foundation of China (No. 52075484). Y.L., C.X. and H.H. contributed equally to this work.

## Conflict of Interest

The authors declare no competing interests.

## Author contributions

Y.L. and B.H. conceived and designed the research. C.X., Y.L. and X.P. performed the sample fabrication. Y.L., M.L. and C.X. conducted the material characterization and optical measurements. Y.L., G.L. and M.L. performed the demonstrations of potential applications. H.H. performed the DFT calculation. Y.L., C.X., H.H., D.M., M.W., T.Z. and B.H. analyzed the results and wrote the manuscript. All authors approved the final version of the manuscript.

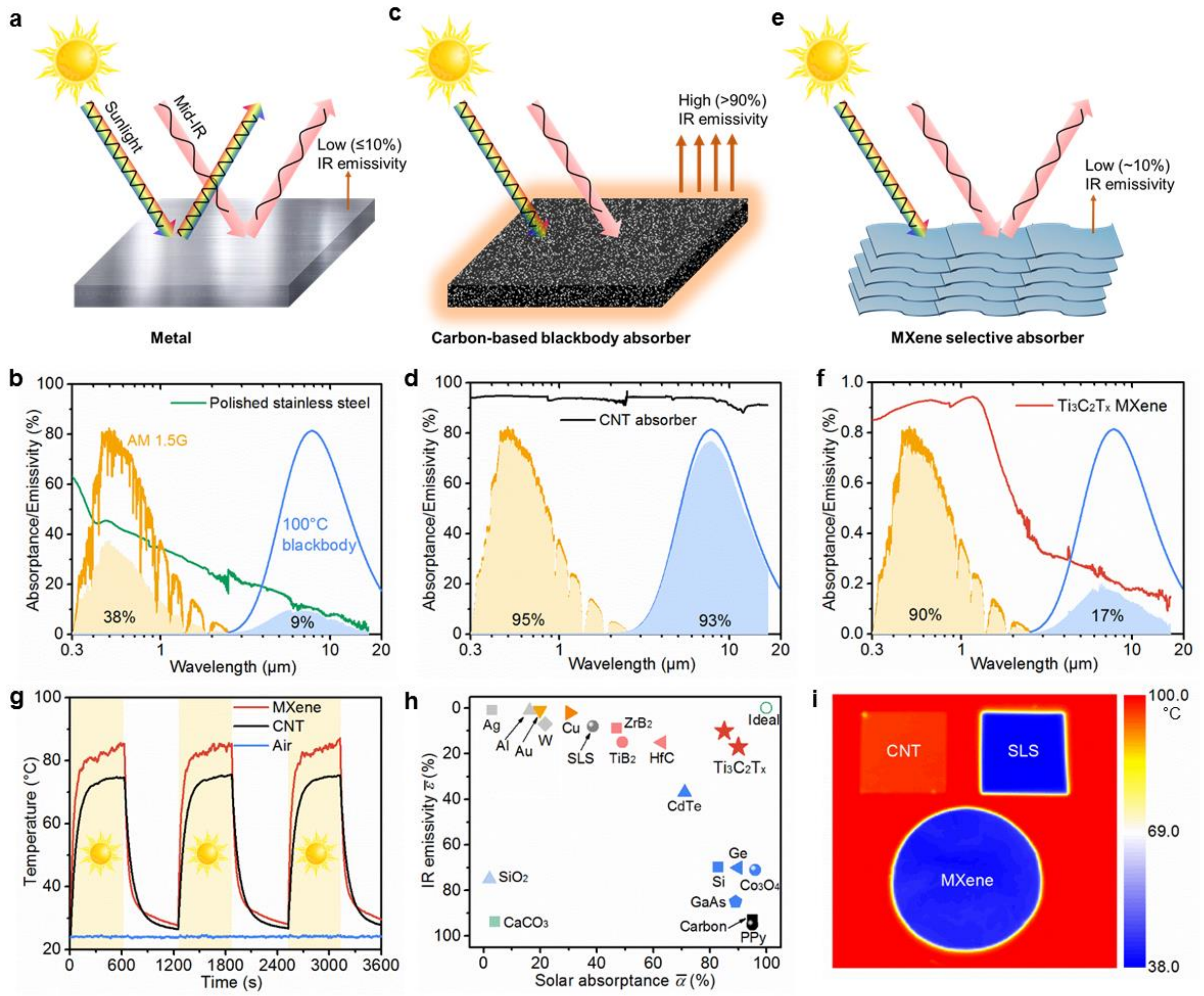

**Figure 1.** Light-matter interaction of metals, carbon-based blackbodies, and MXenes. a) The high solar reflectance, high mid-IR reflectance, and low IR emissivity of metals. c) The high solar absorptance, high mid-IR absorptance, and high IR emissivity of carbon-based blackbodies. e) The high solar absorptance, high mid-IR reflectance, and low IR emissivity of MXenes. b, d, f) Absorptance/emissivity spectra of a polished stainless steel (SLS) sheet, a CNT black absorber, and a $Ti_3C_2T_x$ MXene film, as well as the AM 1.5G solar spectrum and the radiation spectrum of a blackbody at 100 °C. g) Temperature *vs.* time of the CNT and MXene absorbers, and the air temperature under 1 sun. h) Comparison of solar absorptance and IR emissivity of intrinsic materials including metals (Au, Ag, Al, Cu, W, and SLS), radiative cooler materials ($SiO_2$ and $CaCO_3$), semiconductors (Si, Ge, CdTe, GaAs, and $Co_3O_4$), black materials (carbon-based and polymers), $TiB_2$, $ZrB_2$, HfC, and $Ti_3C_2T_x$ MXenes. i) IR photographs of the SLS, CNT, and MXene on a hot plate with a constant temperature of 100 °C.

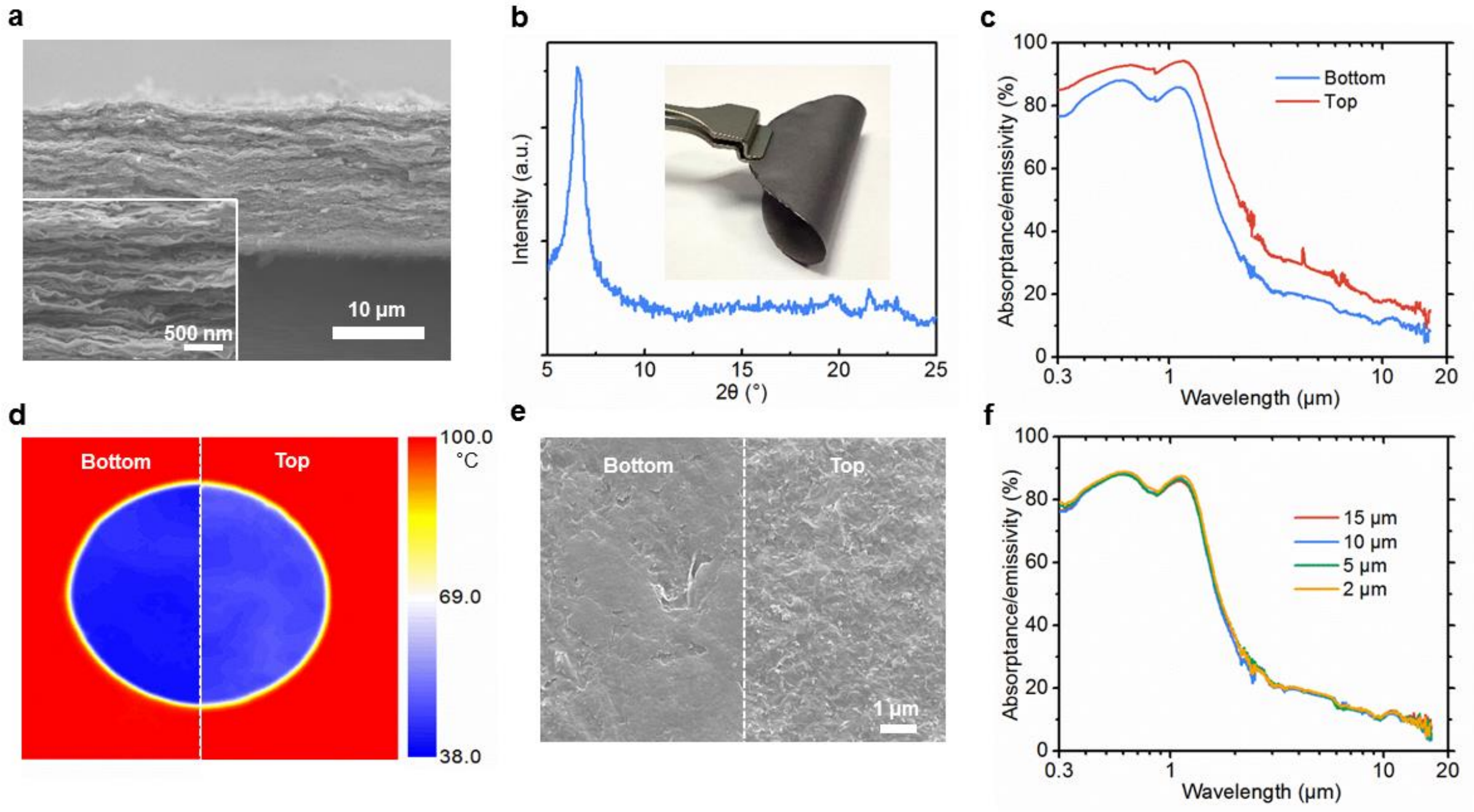

**Figure 2.** Characterizations of the free-standing $Ti_3C_2T_x$ MXene film fabricated by vacuum-assisted filtration. a) Cross-sectional SEM image of the free-standing film. b) XRD pattern of the film. Inset: Photograph of the flexible black film (top side). c) Absorptance/emissivity spectra of the bottom (attached to the filter membrane) and top sides of the 15-µm-thick vacuum-filtrated film. d) IR photographs of the two sides placed on a hot plate (100 °C). e) Surface SEM images of the two sides. f) Absorptance/emissivity spectra of vacuum-filtrated films (bottom sides) with different thicknesses.

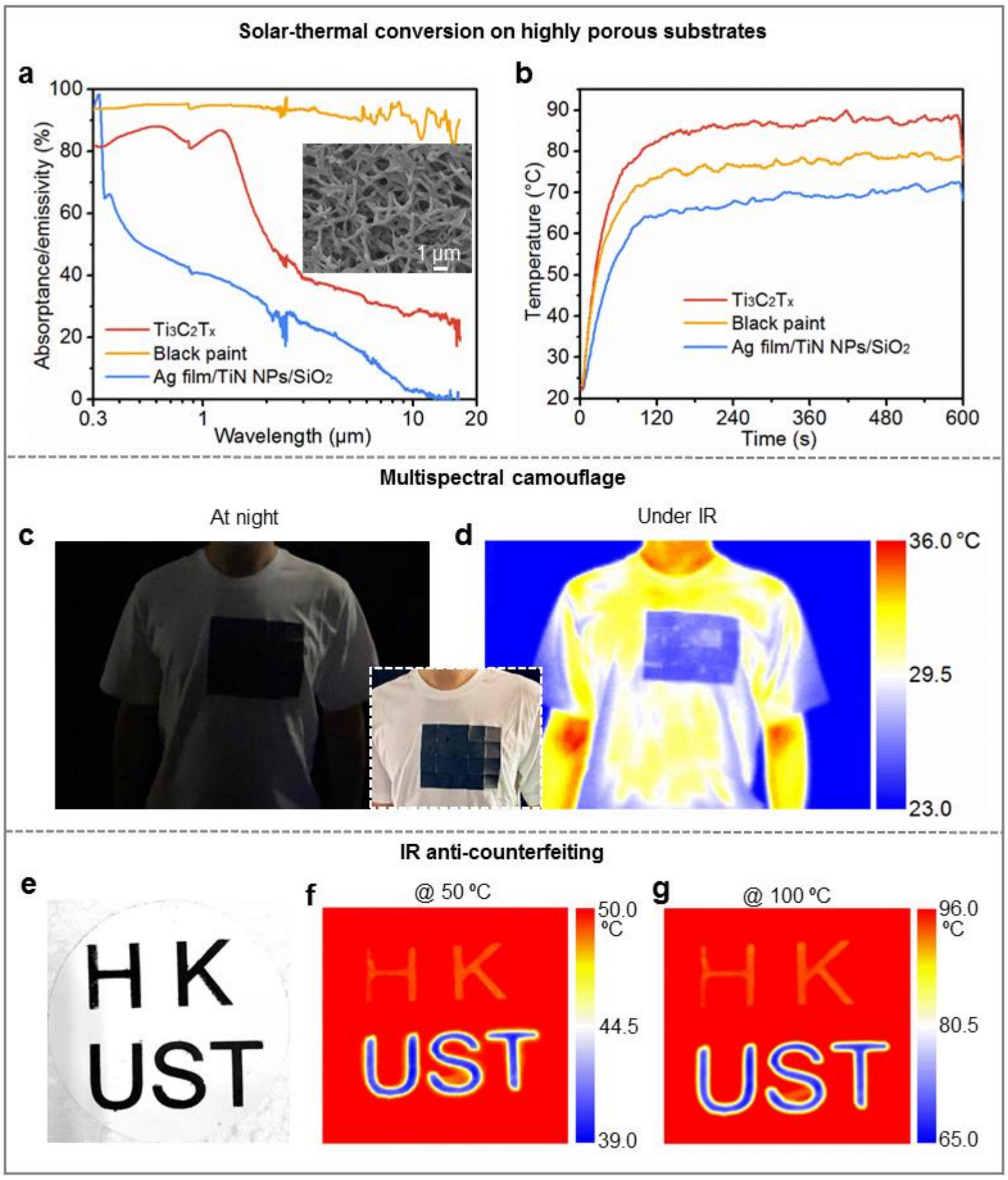

**Figure 3.** Potential applications of the low-emissivity black $Ti_3C_2T_x$. a) Absorptance/emissivity spectra of highly porous Nylon 66 membranes coated with $Ti_3C_2T_x$ and metamaterial-based selective absorbers (Ag/TiN nanoparticles/$SiO_2$). Inset: SEM image of the highly porous nylon membrane. b) Temperature *vs.* time of Nylon 66 membranes coated with $Ti_3C_2T_x$ and metamaterial absorbers under 1 sun. c-d) Optical and IR photographs of a person who wore a white T-shirt with the black $Ti_3C_2T_x$ coating at night. Inset: An optical photograph at daytime. e) Optical photograph of the word "HKUST" written using commercial black paint (for "HK") and $Ti_3C_2T_x$ MXene (for "UST") solutions on a PVDF membrane. f-g) IR photographs of the black word "HKUST" on hot plates with set temperatures of 50 and 100 °C, respectively.

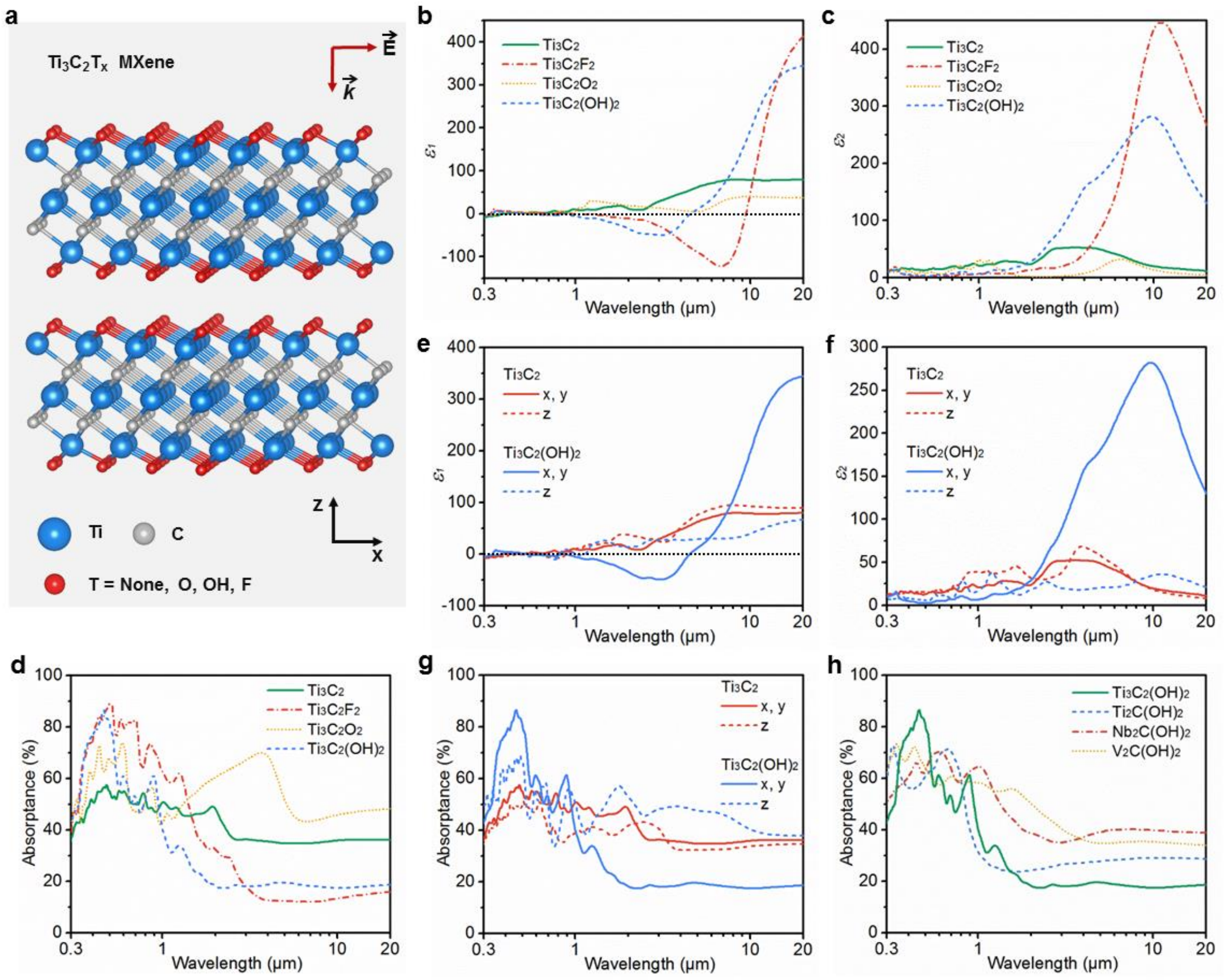

**Figure 4.** First-principles calculations. a) Molecular layered structure of the $Ti_3C_2T_x$ bulk film. b-c) Real and imaginary parts of in-plane permittivity of the $Ti_3C_2T_x$ with different groups. d) Calculated absorptance spectra of the $Ti_3C_2T_x$ with different groups using the in-plane permittivity. e-f) Real and imaginary parts of permittivity of the $Ti_3C_2$ and $Ti_3C_2(OH)_2$ along different directions. g) Calculated absorptance spectra of the $Ti_3C_2$ and $Ti_3C_2(OH)_2$ using permittivity along different directions. h) Calculated absorptance spectra of different MXenes with -OH groups including $Ti_3C_2(OH)_2$, $Ti_2C(OH)_2$, $Nb_2C(OH)_2$, and $V_2C(OH)_2$, using the in-plane permittivity.

# Supporting Information

**2D MXenes: Visible Black but Infrared White Materials**

Yang Li, Cheng Xiong, He Huang, Xudong Peng, Deqing Mei, Meng Li, Gongze Liu, Maochun Wu, Tianshou Zhao,* and Baoling Huang*

**Theoretical calculation**

All first-principles calculations were performed in the framework of the density functional theory (DFT) with the projector augmented wave method as implemented in the Vienna *ab initio* simulation package (VASP).[1] The electronic exchange-correlation interaction was described by the scheme of the generalized gradient approximation (GGA) with the Perdew-Burke-Ernzerhof (PBE) functional.[2,3] All layered bulk MXene crystals including the lattice parameters and atomic positions were allowed to relax until the total energy and residue Hellman-Feynman force were converged to $10^{-6}$ eV and 0.01 eV $Å^{-1}$, respectively. The plane-wave cut-off energy was set to 520 eV and the Monkhorst–Pack k-point meshes of 21×21×5, 41×41×7 and 49×49×7 were used for lattice optimization, self-consistence and optical property calculations, respectively. Empirical DFT-D3 dispersion method [4] was used to consider the interlayer van der Waals interactions. The optical properties of layered Mxene were described by the dielectric function, $\varepsilon(\omega)=\varepsilon_1(\omega)+i\varepsilon_2(\omega)$. The imaginary part $\varepsilon_2(\omega)$ of the dielectric tensor was directly calculated by a summation over the empty states using the equation,[5]

$$\varepsilon_2(\omega)=\frac{2\pi e^2}{\varepsilon_0\Omega}\sum_{c,v,\boldsymbol{k}}\left|\left\langle\Psi_{\boldsymbol{k}}^c\left|\boldsymbol{u}\cdot\boldsymbol{r}\right|\Psi_{\boldsymbol{k}}^v\right\rangle\right|^2\cdot\delta\left(E_{\boldsymbol{k}}^c-E_{\boldsymbol{k}}^v-\hbar\omega\right) \tag{1}$$

where $\Psi_{\boldsymbol{k}}^c$ and $\Psi_{\boldsymbol{k}}^v$ are the wave functions of the conduction band ($c$) and valence band ($v$) at the k-point $\boldsymbol{k}$, respectively, $\boldsymbol{u}$ is the incident electric field polarization, $\boldsymbol{u}\cdot\boldsymbol{r}$ is the momentum operator, $\hbar\omega$ is the incident photon energy, $\varepsilon_0$ is the vacuum dielectric constant and $\Omega$ is the simulation volume. Once the imaginary part was obtained, the real part $\varepsilon_1(\omega)$ could be derived from Kramers-Kronig transformation relation, and the reflectivity and absorption coefficients were estimated by using the standard optical relationship.

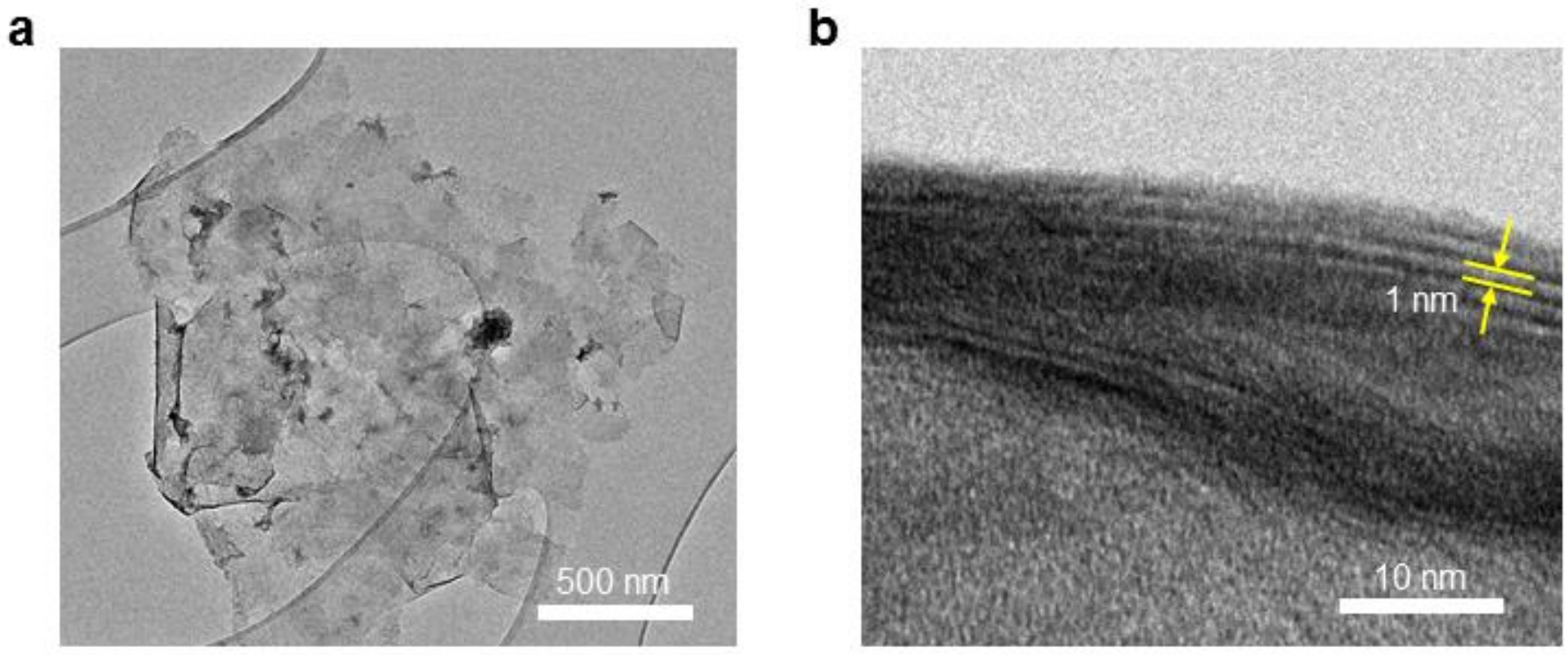

**Figure S1.** TEM images of the $Ti_3C_2T_x$ nanoflakes. a) Low-resolution TEM images of a single $Ti_3C_2T_x$ nanoflake of few layers. b) High-resolution TEM image of the rolling-up nanoflake edge. The layer thickness is around 1 nm.

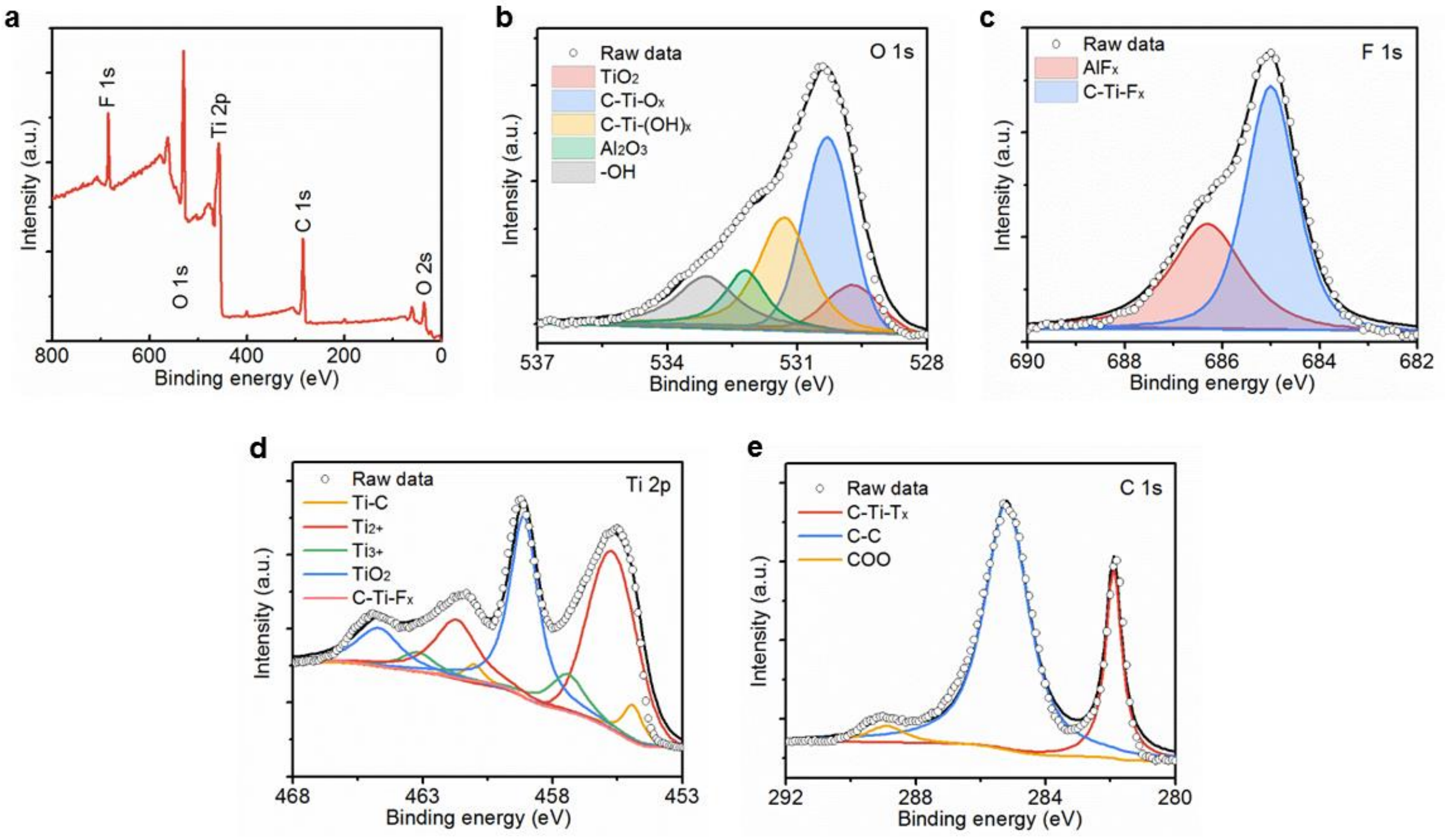

**Figure S2.** XPS spectrum of the film. a) XPS spectrum of the $Ti_3C_2T_x$ film. b-e) High-resolution XPS spectra of O 1s, F 1s, Ti 2p, and C 1s of the film.

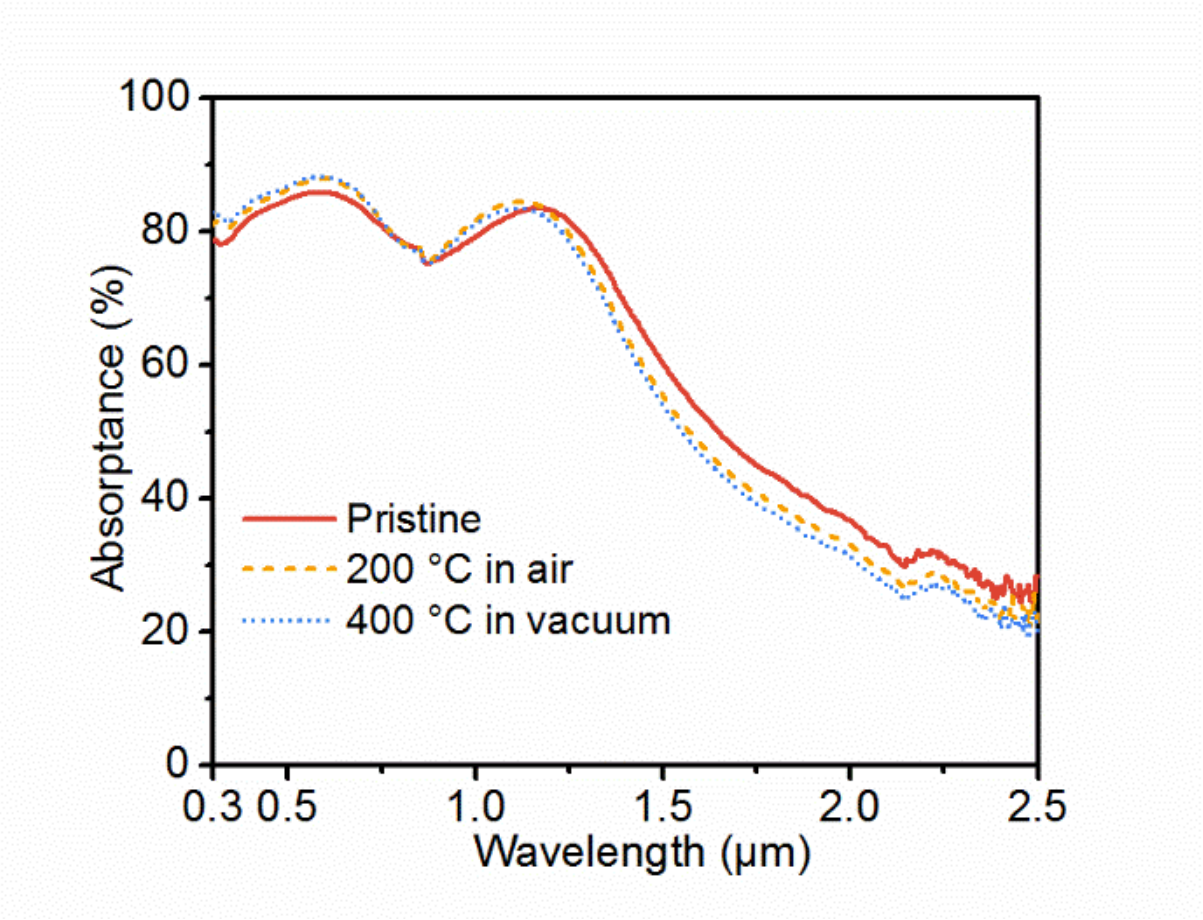

**Figure S3.** Long-term thermal stability. Absorptance spectra of the $Ti_3C_2T_x$ MXene film (bottom side) before and after annealing at 200 °C in the air, and at 400 °C in the vacuum.

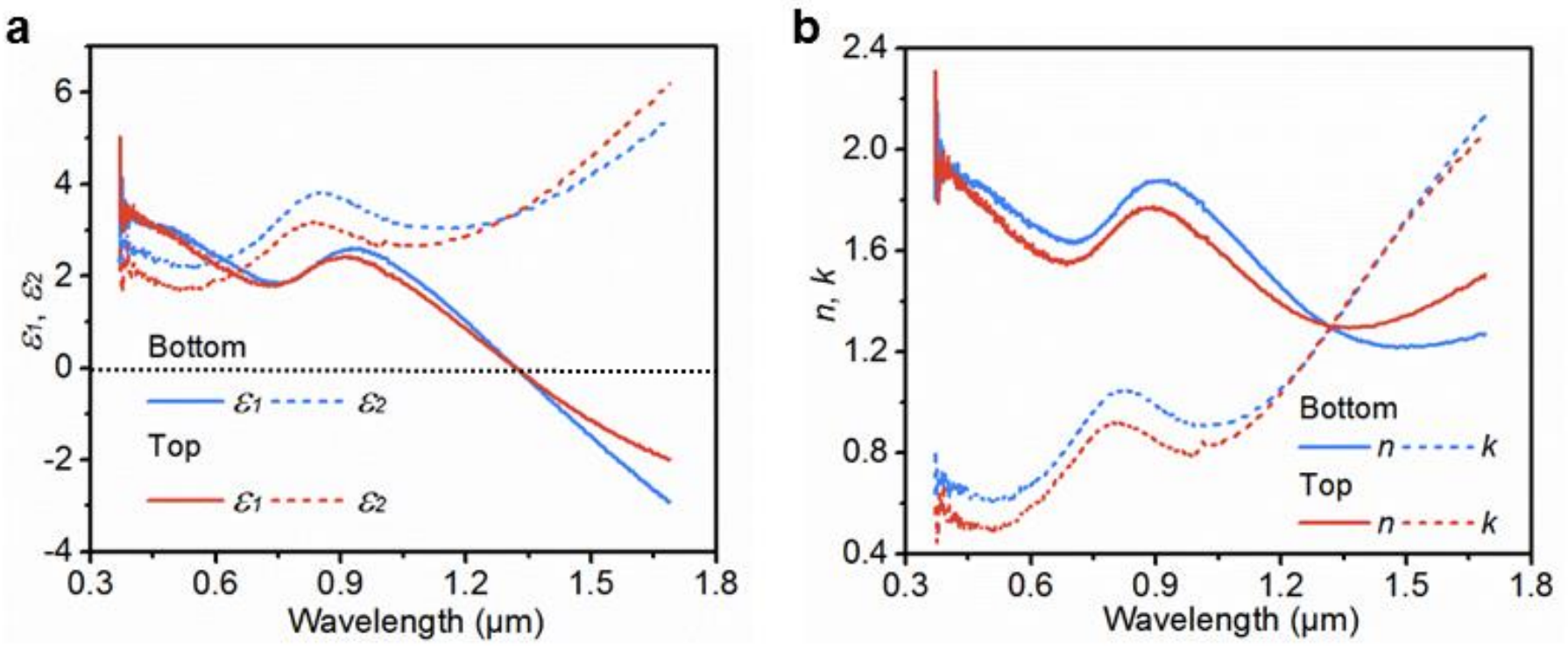

**Figure S4.** Measured optical properties of the two sides of vacuum-filtrated $Ti_3C_2T_x$ films. a) The real part and the imaginary part of permittivity. b) The real part and the imaginary part of refractive index.

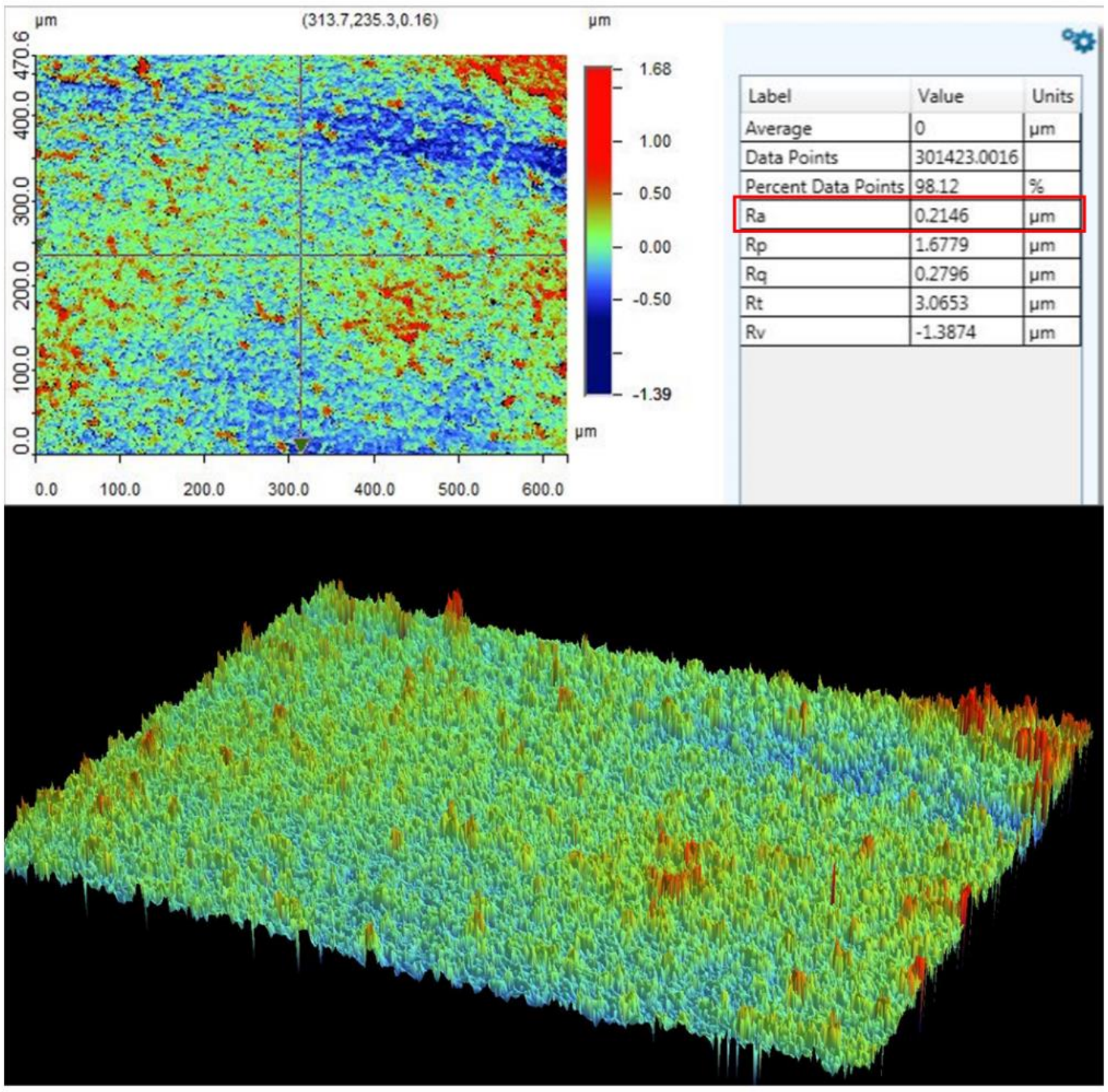

**Figure S5.** 2D and 3D surface profiles of the bottom side of $Ti_3C_2T_x$ films.

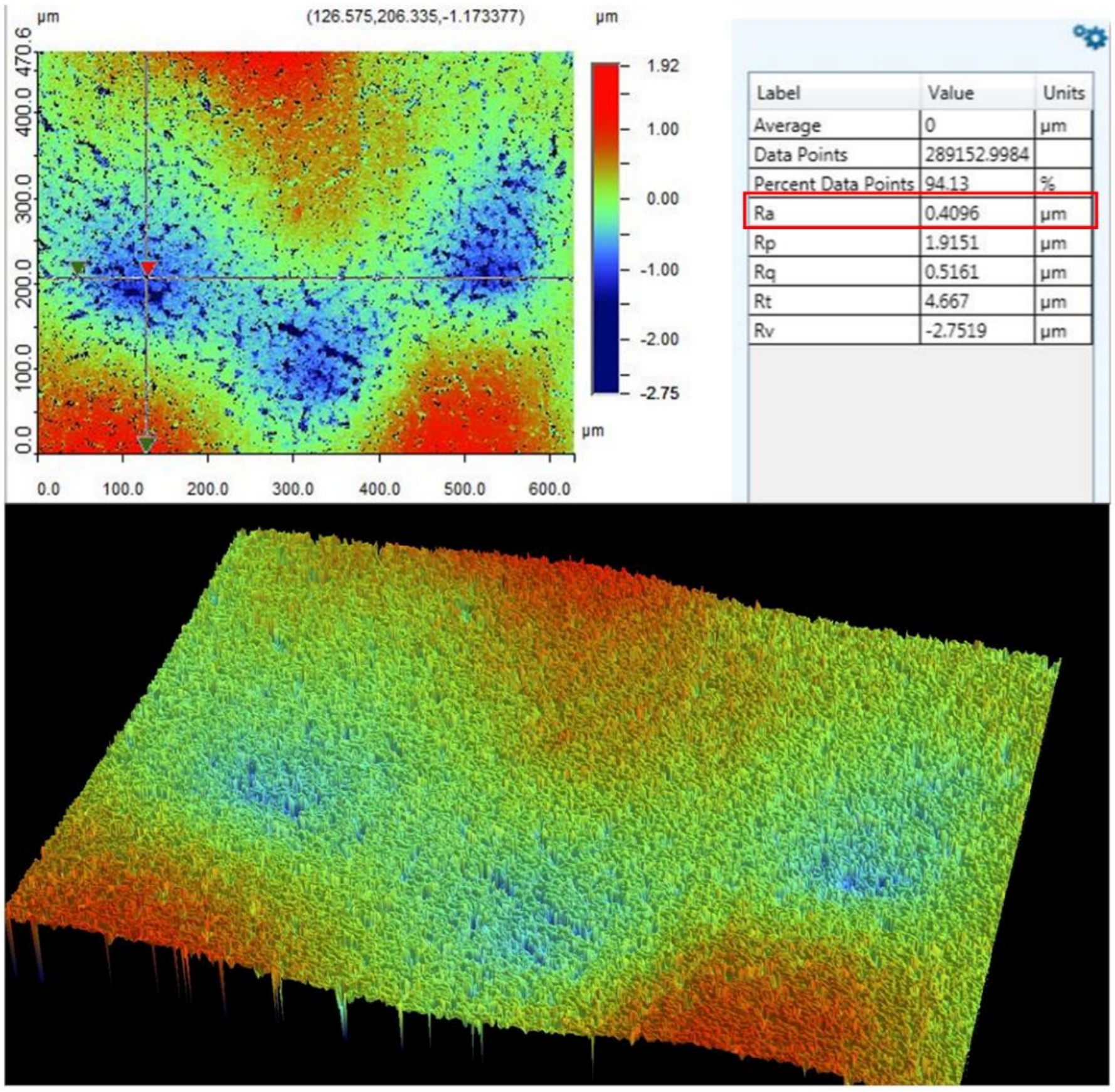

**Figure S6.** 2D and 3D surface profiles of the top side of $Ti_3C_2T_x$ films.

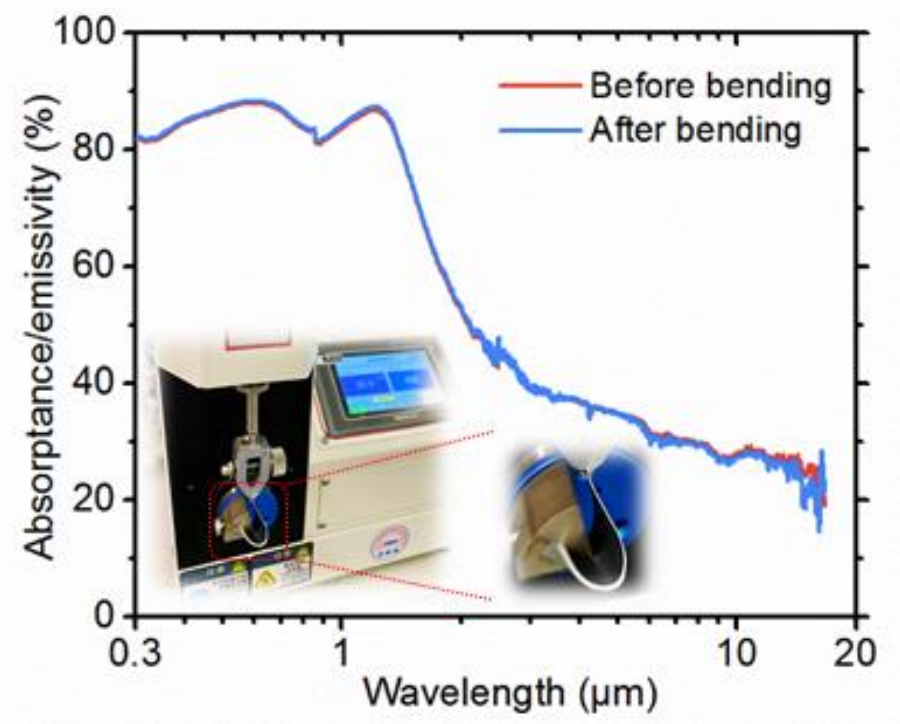

**Figure S7.** Bending test. Absorptance/emissivity spectra of the $Ti_3C_2T_x$-Nylon 66 absorber before and after bending 150 ° for 10,000 time.

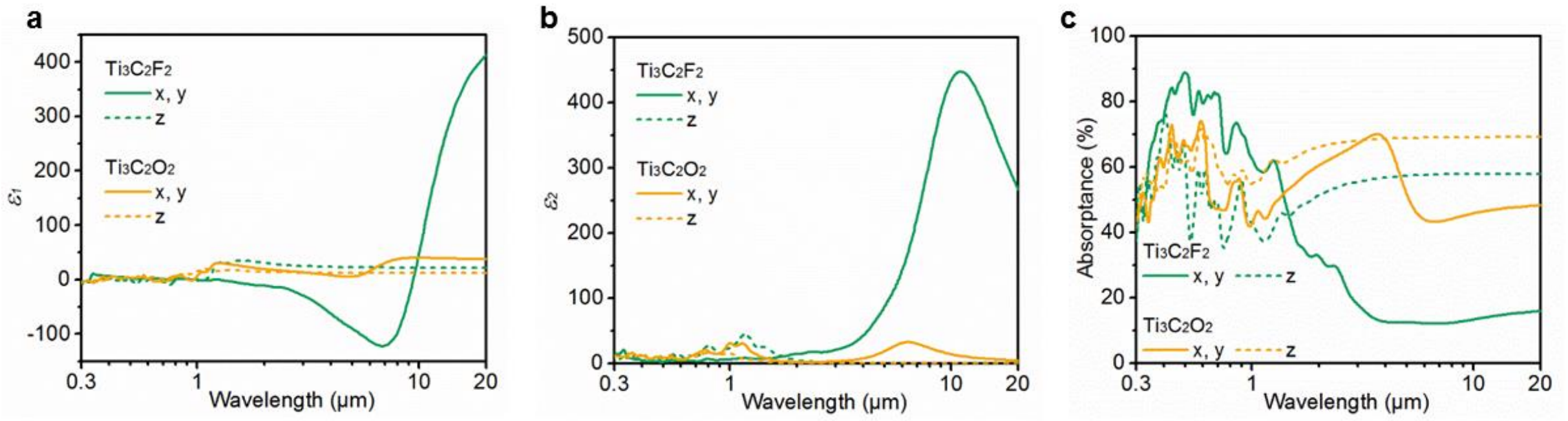

**Figure S8.** Calculated permittivity and UV-visible-IR absorptance spectra of both the $Ti_3C_2F_2$ and $Ti_3C_2O_2$ using DFT. a) Real parts and b) imaginary parts of the permittivity of both $Ti_3C_2F_2$ and $Ti_3C_2O_2$ along different directions. c) The resulting absorptance spectra of both $Ti_3C_2F_2$ and $Ti_3C_2O_2$ along different directions.

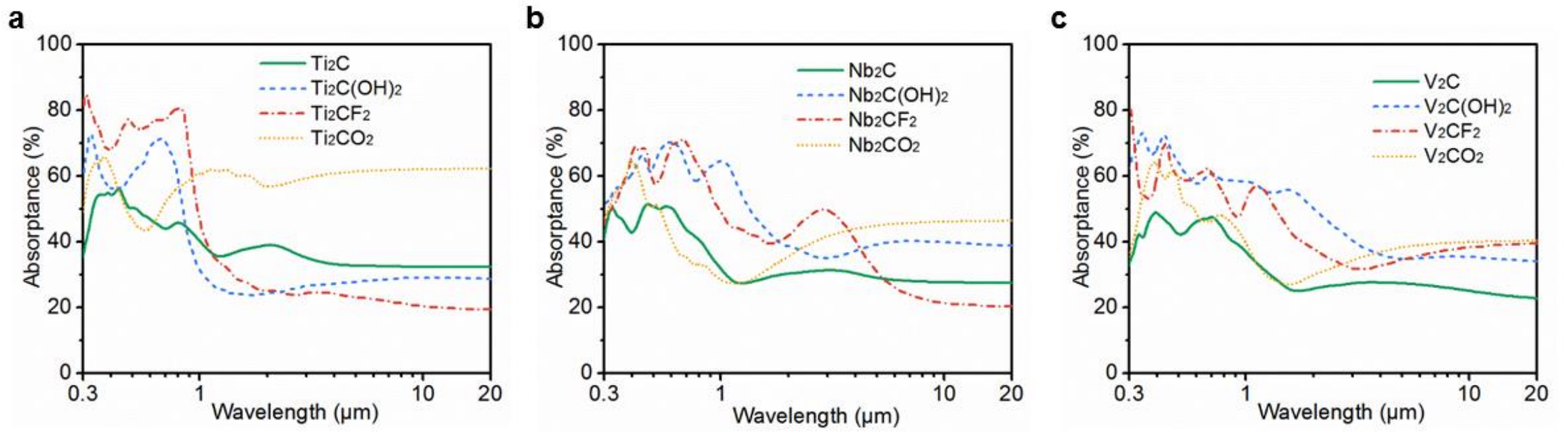

**Figure S9.** Calculated UV-visible-IR absorptance spectra of the $Ti_2CT_x$, $Nb_2CT_x$, and $V_2CT_x$ using DFT. a) Absorptance spectra of the $Ti_2C$, $Ti_2C(OH)_2$, $Ti_2CF_2$, and $Ti_2CO_2$. b) Absorptance spectra of the $Nb_2C$, $Nb_2C(OH)_2$, $Nb_2CF_2$, and $Nb_2CO_2$. c) Absorptance spectra of the $V_2C$, $V_2C(OH)_2$, $V_2CF_2$, and $V_2CO_2$.

**Table S1. Comparison of the optical performance of intrinsic solar absorbing materials.**

| Bulk materials | Solar absorptance $\bar{\alpha}$ | IR emissivity $\bar{\varepsilon}$ | Spectral selectivity $\bar{\alpha}/\bar{\varepsilon}$ |
|---|---|---|---|
| W | 22% | 7% | 3.1 |
| Stainless steel | 38% | 9% | 4.2 |
| Si [6] | 83% | 70% | 1.2 |
| Ge [6] | 90% | 70% | 1.3 |
| $Co_3O_4$ [6] | 96% | 71% | 1.4 |
| $SiO_2$ | 2% | 75% | - |
| $ZrB_2$ [7] | 47% | 9% | 5.2 |
| $TiB_2$ [7] | 49% | 15% | 3.3 |
| HfC [8] | 63% | 15% | 4.2 |
| SiC | 95% | 90% | 1.1 |
| ITO [9] | - | 19% | - |
| CdTe [10] | 71% | 37% | 1.9 |
| GaAs [11] | 89% | 85% | 1.1 |
| Carbon-based | ~95% | 90-95% | ~1 |
| **$Ti_3C_2T_x$ MXene** | **82% (90%)** | **10% (17%)** | **8.2 (5.3)** |